# Laboratory observation of ion acceleration via reflection off laser-produced magnetized collisionless shocks

Hui-bo Tang[1,2,†], Yu-fei, Hao[1,3,6, †], Guang-yue Hu[1,4,5,*], Quan-ming Lu[1,2,6,*], Chuang Ren[7], Yu Zhang[7], Ao Guo[1, 2], Peng Hu[1,4], Yu-lin Wang[1,4], Xiang-bing Wang[1,4], Zhen-chi Zhang[1,4], Peng Yuan[1,4], Wei Liu[1,4], Hua-chong Si[1,4], Chun-kai Yu[1,4], Jia-yi Zhao[1,4], Jin-can Wang[1,4], Zhe Zhang[8], Xiao-hui Yuan[9], Da-wei Yuan[10], Zhi-yong Xie[11], Jun Xiong[11], Zhi-heng Fang[11], Jian-cai Xu[5], Jing-Jing Ju[5], Guo-qiang, Zhang[13], Jian-Qiang Zhu[12], Bai-fei Shen[5], Ru-xin Li[5], Zhi-zhan Xu[5]

[1]CAS Key Laboratory of Geospace Environment, University of Science and Technology of China, Hefei, China

[2]School of Earth and Space Sciences, University of Science and Technology of China, Hefei, China

[3]Key Laboratory of Planetary Sciences, Purple Mountain Observatory, Chinese Academy of Sciences, Nanjing, China

[4]School of Nuclear Science and Technology & School of Physical Science, University of Science and Technology of China, Hefei, China

[5]State Key Laboratory of High Field Laser Physics & CAS Center for Excellence in Ultra-intense Laser Science, Shanghai Institute of Optics and Fine Mechanics, Chinese Academy of Sciences, Shanghai, China,

[6]CAS Center for Excellence in Comparative Planetology, Hefei, China

[7]Department of Mechanical Engineering, University of Rochester, Rochester, New York, USA

[8]Institute of Physics, Chinese Academy of Sciences, Beijing, China

[9]Key Laboratory for Laser Plasmas (Ministry of Education), School of Physics and Astronomy, Shanghai Jiao Tong University, Shanghai, China

[10]Key Laboratory of Optical Astronomy, National Astronomical Observatories, Chinese Academy of Sciences, Beijing, China

[11]Shanghai Institute of Laser Plasma, Shanghai, China

[12]National Laboratory on High Power Laser and Physics, Shanghai Institute of Optics and Fine Mechanics, Chinese Academy of Sciences, Shanghai, China

[13]Shanghai Institute of Applied Physics, Chinese Academy of Sciences, Shanghai, China

[†]These authors contributed equally: Huibo Tang, Yufei Hao.

*Corresponding authors: Guang-yue Hu; Quanming Lu

*Email: gyhu@ustc.edu.cn; qmlu@ustc.edu.cn

Fermi acceleration by collisionless shocks is believed to be the primary mechanism to produce high-energy charged particles in the Universe, where charged particles gain energy successively from multiple reflections off the shock front. Here, we present the first direct laboratory experimental evidence of ion energization from single reflection off a supercritical quasi-perpendicular collisionless shock, an essential component of Fermi acceleration, in a laser-produced magnetized plasma. We observed a quasi-monoenergetic ion beam with 2-4 times the shock velocity in the upstream flow using time-of-flight method. Our related kinetic simulations reproduced the energy gain and showed that these ions were accelerated mainly by the motional electric field during reflection off the shock front. Our experimental results are consistent with the quasi-monoenergetic fast ion component observed in the Earth's bow shock, and open the way for controlled laboratory investigations of the cosmic accelerators.

Collisionless shocks are among the most powerful particle accelerators in astrophysics [1,2]. They act as the moving scattering centers, originally proposed by Fermi as an origin of cosmic rays [3], where charged particles gain energy by reflecting off them. A succession of small energy increments due to repeated shock crossings back and forth between the upstream and downstream creates the power law spectrum of energetic particles, a process known as diffusive shock acceleration (DSA) [4-8]. To enter the Fermi energization cycle, particles must be pre-accelerated to have a gyroradius large enough to be able to scatter between upstream and downstream. Several competing mechanisms have been proposed to solve this well-known 'injection problem' [9-11], all in theory or simulations [12-25].

Substantial efforts have been devoted to measure the formation and charged particle energization of collisionless shock in laboratory experiments in the past decade [26-34], but the ion energization from single reflection off the shock front, a key step in Fermi shock acceleration and also a potential injection mechanism [9-11], has never been observed in laboratory magnetized shock experiments [35-39] until this work.

Here, we report on experimental results of ion acceleration in a supercritical quasi-perpendicular collisionless shock formed when a laser-produced supersonic plasma flow impact on a magnetized ambient plasma. Quasi-monoenergetic ions with 2-4 times the shock velocity are observed in the upstream of shock, which are produced mainly by the motional electric field acceleration during specular reflection from the shock. It's the first direct laboratory experimental evidence of ion acceleration from single reflection off a collisionless shock. The experimental feature of quasi-monoenergetic ion distribution is in well agreement with the fast ion component observed in the Earth's bow shock.

The experiments were conducted at the Shenguang II (SG II) laser facility. A sketch of the experimental setup is shown in Fig. 1a. A weaker precursor laser beam (~1×10$^{13}$ W/cm$^2$) ablated a plastic (CH$_2$) planar target to create the ambient plasma, which was magnetized by a 4-6 T external background magnetic field [40] via an anomalously fast magnetic diffusion process [37,41-43]. An intense drive laser beam (~8×10$^{13}$ W/cm$^2$) irradiated another plastic (CH$_2$) target with a focus spot of 0.5 × 0.5 mm$^2$ to produce supersonic plasma flow as a piston. The piston plasma flow drove a quasi-perpendicular collisionless shock in the magnetized ambient plasma. The profile of the shock and the ambient plasma density were characterized with optical diagnostics. The ion velocity spectrum was measured by the time-of-flight method using a Faraday Cup (see Methods for further details).

The electron density of the ambient plasma varies from ~1×10$^{18}$/cm3 to 5×10$^{18}$/cm$^3$ with a gradient scale length of ~1 mm (Fig. 1c), and the electron temperature is estimated to be ~40±10 eV [37,44]. The piston plasma with a higher electron temperature of ~200 eV [37,44] drives a quasi-hemispherical magnetized collisionless shock (Fig. 1b), which is asymmetric due to the inhomogeneity of the ambient plasma (Fig. 1c). The angle between the shock normal and the upstream magnetic field $\theta_{Bn}$ in our experiments is approximately 90°; therefore, it is a nearly perpendicular shock (see Methods for further details). The shock velocity is $v_{shock}$~400 km/s over the span of measurement, which is slightly slower than that without an external magnetic field (Supplementary Fig. S4), yet still within the measurement error. A strongly compressed zone is formed within the plasma, which exhibits typical structures of a "foot", a "ramp", et al (Fig. 1d, Supplementary Fig.S3) [36, 38, 39, 45].

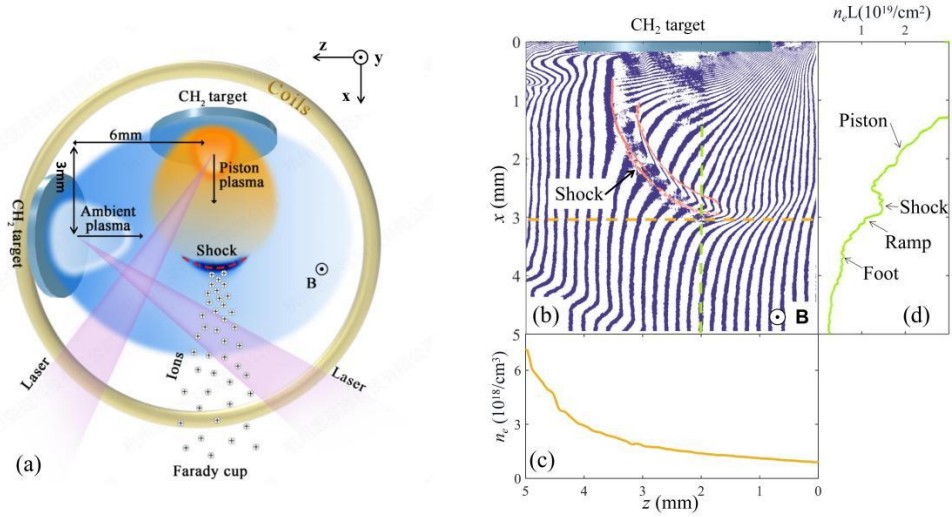

**Fig. 1| Laser-driven magnetized collisionless shock experiments. a**, Sketch of the experimental setup: A 4-6 T external magnetic field (along the y direction) was applied by a pulsing current through a set of magnetic field coils. Ambient plasma was generated after the plastic $CH_2$ target (left) was ablated by a weaker precursor beam (100 J/1 ns/351 nm). After 12 ns (at time $t_0$) that the ambient plasma was magnetized, an intense drive beam (260 J/1 ns/351 nm) irradiated another plastic $CH_2$ target (top) to produce a supersonic piston plasma flow, which drove the collisionless shock in the magnetized ambient plasma. The density profiles of the shock and the ambient plasma were characterized with optical diagnostics. The acceleration of ions was measured by the time-of-flight method using a Faraday Cup (directed along the x-axis). **b**, The imaging of shock measured by optical interferometry (blue) and dark-field schlieren method (red) (line-integrated along y direction), taken at time $t_0 + 4$ ns, formed in the ambient plasma with a 5T external magnetic field. The bright refractive fringes in the optical dark-field schlieren imaging (red), which are the first derivative of the line integrated plasma density, indicate the discontinuity surfaces around the shock. The inhomogeneous ambient plasma results in an asymmetric quasi-hemispherical shock. **c**, Electron density profile for the ambient plasma, taken at time $t_0 + 4$ ns along the yellow line in (b) at x=3 mm, under the experimental condition without a piston plasma flow, which varies from $n_{e0} \sim 1\times10^{18}/cm^3$ to $5\times10^{18}/cm^3$ with a gradient scale length of ~1 mm in the shock traveling zone. **d**, Line-integrated electron density profile of shock taken along the green line in (b) at z=2 mm. L is the plasma size in y direction. The electron densities in upstream and downstream are approximately $1-5\times10^{18}/cm^3$ and $0.5-1.5\times10^{19}/cm^3$ (see details in Supplementary Fig.S3), respectively, which indicate a compression ratio of >3.

Under our experimental parameters, the magnetized shock is approximately collisionless. The ion-ion collisional mean free path is approximately 4 mm, which is much larger than the ion Larmor radius of ~800 μm and the shock thickness of ~500 μm. The >3× density compression factor approximately satisfies the hydrodynamic Rankine-Hugoniot (RH) jump condition of shock [45]. The shock Alfvenic and sonic Mach numbers are $M_A$~7-11 and $M_s$~7-10, respectively, and the ambient plasma beta value is $\beta$~0.3-1.2. Therefore, the shock conditions probed in our experiments are relevant to the Earth's bow shock, where the typical shock Alfvenic Mach number is $M_A$~3-10 [46-52], as illustrated in Table 1.

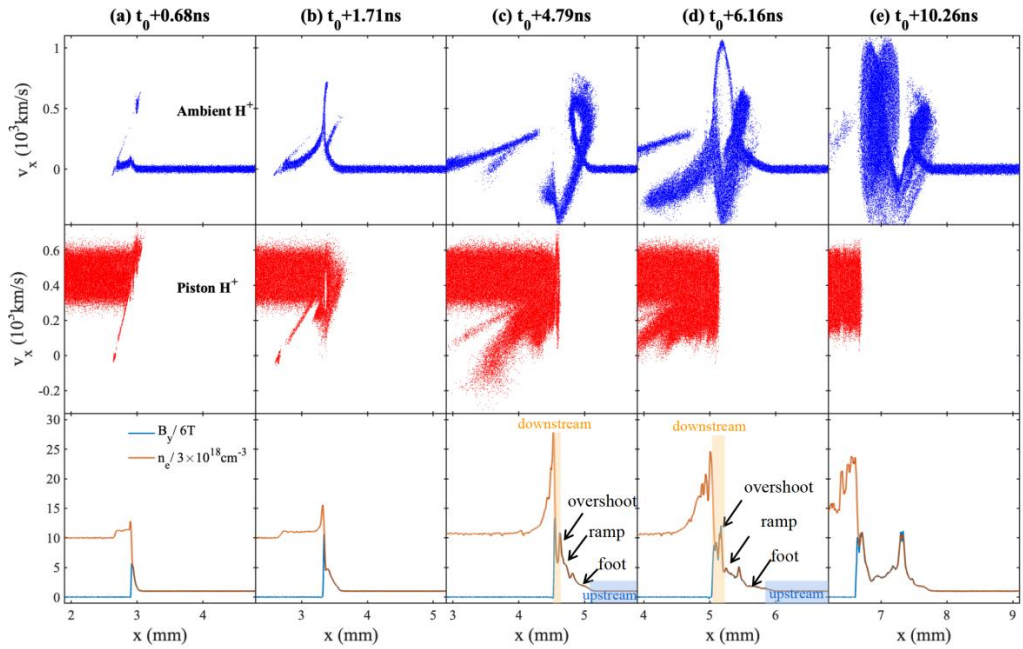

**Fig. 2| Formation of a shock structure and the associated ion dynamics in the 1-D PIC simulation.** The $v_{px}$-$x$ phase space scatter plots of the ambient (blue, **1st row**) and piston (red, **2nd row**) $H^+$ ions to present the ion dynamics associated with shock formation. (**3rd row**) The magnetic field (blue) and the electron number density (red) profiles are displayed to show the formation of the piston-driven shock. The time steps of $t_0$+0.68 ns (**a**), $t_0$+1.71 ns (**b**) and $t_0$+4.79 ns (**c**) correspond to the early time before shock formation, onset of shock formation (~$\omega_{ci-H}^{-1}$=1.71 ns, which is the upstream $H^+$ ion gyroperiod), and shock formation on ion scales that separated from the piston

($t=t_0+4.79$ ns $> 2\omega_{ci\text{-}H}^{-1}$), respectively. **d**, **e** indicate the shock reformation after distinctly separating from the piston (see details in Supplementary Fig.S8). The proton-to-electron mass ratio is set as $m_p/m_e=100$.

One-dimensional (1-D) and two-dimensional (2-D) particle-in-cell (PIC) simulations are conducted to study the shock formation in piston-driven magnetized ambient plasma under conditions similar to our experimental parameters (see details in Methods), as illustrated in Fig. 2. At the beginning of the interaction, the piston acts like a snowplow with a speed of ~400 km/s and sweeps up the ambient ions and magnetic field (Fig. 2a), which produces density and magnetic field compression around the piston-ambient plasma interface. The particle trajectories indicate that the ions from the ambient and piston plasmas penetrate each other since the ions are effectively collisionless. Within $t_0+1.71$ ns ($\omega_{ci\text{-}H}^{-1}$ ~1.71 ns, the upstream $H^+$ ion gyroperiod), the compressed steepened magnetic structure is strong enough to reflect the ambient $H^+$ ions, at which time the shock begins to form (onset of shock, Fig. 2b) [53]. After distinct separation from the piston, at approximately $t_0+4.79$ ns, a shock on ion scales is formed with a speed of 415 km/s and $M_A$~8.3 (Fig. 2c). Consistent with our experimental results, the shock in the simulation reproduces the characteristic feature of a "foot", a "ramp", and the compression ratio is >3. In the following several gyroperiods, the shock reformation is observed in the shock "foot" region, and the $C^{5+}$ ions form another shock behind the $H^+$ ions shock (Fig. 2d&e, and Supplementary Fig. S8&14).

Ion acceleration is observed in our experiments accompanied by the formation of the magnetized collisionless shock. The time-of-flight signal of ion flux (Fig. 3a), collected along the symmetric axis of the piston flow by the Faraday cup, presents two peaks in the ion velocity spectra (Fig. 3b). The first peak corresponds to the particles coming from the piston plasma, and the velocity is $v_{piston}$~300-800 km/s, which is close to the shock speed ($v_{shock}$~400 km/s). The second peak with the velocity $V_{fast\_ions}$~1100-1800 km/s, generated by the accelerated fast ions, is found to have a quasi-monoenergetic spectrum and is approximately 2-4 times the shock speed, similar to the fast ion component observed in the Earth's bow shock by satellites [46,51,52,54]. We have also changed the strength of the external magnetic field in the experiments and found that the fast ion peak becomes more pronounced with increasing external magnetic field (Fig. 3(b)). Even in the absence of external magnetic field, we still can observe the fast ion peak probably due to the self-generated magnetic field of approximately 1 T [37] (see Supplementary Fig. S5 for further details).

The PIC simulations of the experimental piston-ambient interaction, which also exhibit two peaks in the ion velocity spectra (Fig. 3c), confirming the ion acceleration capability of shock. The first peak of slow ions is provided by the piston plasma downstream of the shock. The second peak is the fast reflected ions upstream with approximately 2-3 times the shock speed. $H^+$ ions picked up from the ambient plasma dominate the fast ions and are accelerated during reflection by the shock (see Supplementary Section IV). Shock formation and ions acceleration are not observed in simulations with approximately zero external magnetic field. Notably, the detailed characteristics of the ion velocity spectra in our simulation cannot be straightforwardly compared with experiments for the following reasons. Firstly, the experiments results are temporally and spatially integrated with ions escaping from the 2-D hemispherical shock with an inhomogeneous background profile, while the simulation is just 1-D or 2-D homogeneous background with reduced proton-to-electron mass ratio to lessen computational burden. Secondly, the magnetized ambient plasma has finite size of ⩽ 10mm in experiments (Fig. 1, and Supplementary Fig. S1). Thus, the reflected ambient ions can escape into vacuum before gyrating back into downstream when the shock reach the boundary of the magnetized ambient plasma, and move ballistically into detector (see Methods and Supplementary Fig.S6 for further details).

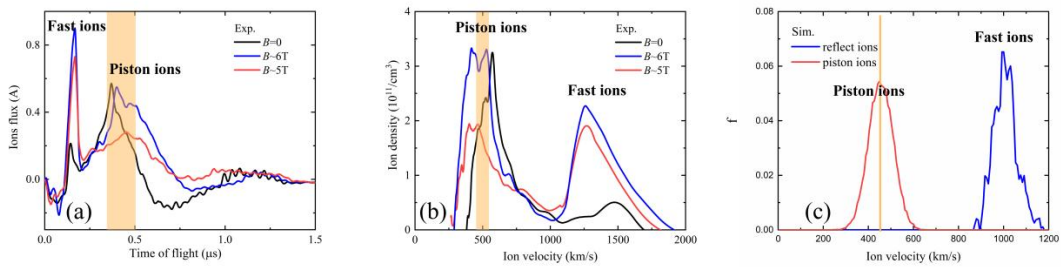

**Fig. 3| Ion velocity spectra in experiments and 1-D PIC simulations. a**, Time-of-flight trace of ion flux in the experiments recorded by Faraday Cup along the symmetric axis of the piston plasma flow. After the precursor negative peak of the noise baseline (0-0.1 μs), the fast ions arrive at the Faraday Cup first at ~0.16 μs, followed by the slow ions (piston) at ~0.4 μs. **b**, Ion velocity spectra in the experiments that transform the time-of-flight trace of ion flux (shown in (a)) to the collected ion density profile in a Faraday cup (see methods and Supplementary Fig. S5). The slow ions with velocity $v$~300-700 km/s come from the piston plasma. The fast ions with velocity $v$~1100-1800 km/s, with approximately 2-4 times the shock speed, are the population from ambient ions accelerated by the shock. **c**, Ion velocity spectrum collected in the foot region of the shock (x> 8mm region at $t_0$+11ns, Supplementary Fig. S8) from the simulation with an external magnetic field of 6 T, which also exhibits two peaks. The velocity of the slow ions is ~400 km/s, while that of the fast ions is ~900-1200 km/s. The shock position is indicated by the orange shaded region.

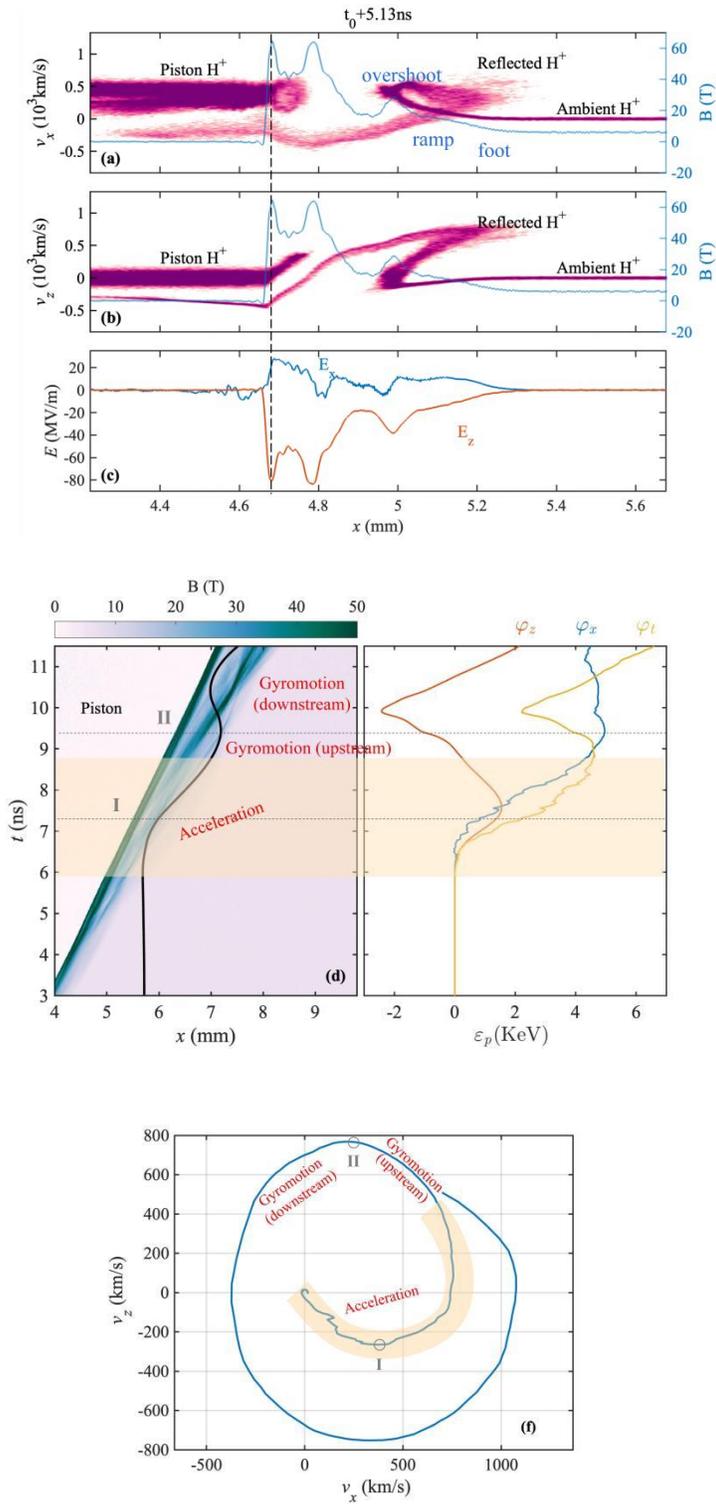

**Fig. 4| Ion acceleration in 1-D PIC simulations. a, b,** The $v_{px}$-$x$ (a) and $v_{pz}$-$x$ (b) phase space scatter plots of the H$^+$ ions at $t_0$+5.13 ns (normalized, including ambient and piston plasma), along

with the profile of the magnetic field (blue line). **c**, The $E_x$ (blue) and $E_z$ (red) electric fields at $t_0$+5.05 ns. **d**, The trajectory (black) of a typical reflected H$^+$ ion originating from ambient plasma overlaid on the profile of the magnetic field strength (color bar). **e**, The time history of the potential gain of the reflected H$^+$ ion $\varphi_x$ (olive), $\varphi_z$ (pink), and $\varphi_t$ (black) ( $\varphi_i = \int_t E_i v_i dt, i = x, z$ , and the total potential gain $\varphi_t = \varphi_x + \varphi_z$ ). **f**, The H$^+$ ion trajectory in $v_z - v_x$ space. The external magnetic field $B_y$ is 6T. The interface between shock and piston is labeled approximately with dash line in **a-c**. In **d-f**, the reflection and acceleration stage is indicated by the orange shaded region, while the moments of ion reflection and that ion gyrates back into downstream are labeled with lines/circles I and II, respectively.

As indicated in our simulations, there exist two components of electric fields $E_x$ and $E_z$ associated with the shock (Fig. 4c). The electric field $E_x$ is an electrostatic field caused by motional electric field and charge separation, while the electric field $E_z$ is only a motional electric field [53,55] (~$v_{shock}B_d$, where $B_d$ is the magnetic field downstream). Our simulations indicate that 99.9% of the accelerated ions experience single reflection, and more than 73% of them undergo shock drift acceleration (Supplementary Fig. S10&11). Most of the accelerated ions are H$^+$, and C$^{5+}$ ions ratio is less than 1%. The reflection efficiency of the ambient ions is about 20%-26% in 1-D and 2-D simulation. By following the trajectory of a randomly chosen typical single reflected drift accelerated H$^+$ ion described in Fig. 4d-e, we can identify the particle energization around the shock, which is dominated by the motional electric field (Supplementary Fig. S9), can be approximately separated into two stages. In the first stage of "reflection and acceleration" (the orange shaded region in fig.4d), the H$^+$ ion slides into the shock "foot" (~6.0 ns), and gets accelerated by the $E_x$ and $E_z$ field. At ~ 7.2 ns, the H$^+$ ion is reflected toward upstream, followed by further acceleration until escape from shock transition layer into upstream region. Then the reflected H$^+$ ion starts the second stage of "gyromotion" at ~8.7 ns in upstream region with little energization. Subsequent to energization, part of the reflected H$^+$ ions gyrate into the downstream region and dissipate energy in it, while the remaining H$^+$ ions are still in the upstream region, which can escape into vacuum when the shock moves to the boundary of the magnetized ambient plasma of finite size (Supplementary Fig. S6), and produces the quasi-monoenergetic fast ion peak collected by the Faraday cup in our experiments. Assuming that the acceleration timescale in the motional electric field is approximately one gyroperiod $m/(qB_{ave})$ ($B_{ave}$ is the average magnetic field that the reflected ions are experienced around the shock), the velocity gain of the reflected ions in the $z$ direction can be estimated as $\Delta v_z \sim v_{shock}B_d/B_{ave} \sim (1\text{-}3)v_{shock}$. Therefore, the reflected ions have a speed of approximately $v \sim \sqrt{\Delta v_x^2 + \Delta v_z^2} \sim (1.4\text{-}3.2)v_{shock}$, consistent with our experiments. This mechanism is in well agreement with the satellite observations that a fast ion component exists in the Earth's bow shock with approximate 2 times the shock velocity. We found that a small fraction (<0.1%) of the earlier reflected ions can undergo multiple reflections and acceleration between upstream and shock front, producing higher energy ions with a continuous spectrum that ends up in the downstream region (Supplementary Fig. S12), similar to that has been observed recently [39], and potentially start the Fermi energization cycle. While the higher-energy ions are 3-4 orders of magnitude weaker than the quasi-monoenergetic ion peak in our experiments, thus it will be hidden under our experimental noise baseline.

In conclusion, our results provide the first direct laboratory experimental evidence of ion energization from single reflection off a supercritical quasi-perpendicular collisionless shock, which are consistent with the satellite observations of the quasi-monoenergetic fast ion component in the Earth's bow shock [46,51,52,54]. Repeated reflections from collisionless shock, accompanied by successive small energy increments, have the potential to push charged-particle energies up to very high values for initiating the Fermi acceleration cycle and producing the high-energy charged particles in the Universe. This opens the path for controlled laboratory experiments that can greatly complement remote sensing and spacecraft observations and help validate particle acceleration models.

| Parameters | Our exp. | Our sim. | Bow shock[47-52] | Term. Shock[56] | SNR (SN1006)[34] |
|---|---|---|---|---|---|
| Flow velocity (km/s) | 400-500 | 400 | 400 | 300 | 3000-5000 |
| B (G) | $6\times10^4$ | $6\times10^4$ | $6\times10^{-5}$ | $1\times10^{-6}$ | $3\times10^{-6}$ |
| Electron temperature(eV) | 40±10 | 60 | 15 | 1 | 1 |
| Sound velocity $c_s$ (km/s) | 50 | 40 | 50 | 13 | 13 |
| Alfvenic velocity (km/s) | 60 | 50 | 50 | 49 | 15 |
| Ion Thermal Velocity (km/s) | 130 | 22 | 50 | 10 | 10 |
| Collisional mean free path $\lambda_{mfp}$ (cm) | 0.4 | | $1\times10^{16}$ | $1.3\times10^{19}$ | $3\times10^{21}$ |
| Ion Larmor Radius $r_{ci}$ (cm) | 0.08 | | $7\times10^6$ | $1\times10^8$ | $3.4\times10^7$ |
| $\lambda_{mfp}/r_{ci}$ | 5 | | $2\times10^9$ | $1\times10^{11}$ | $1\times10^{14}$ |
| beta | 0.3-1.2 | 1.005 | 1.2 | 0.081 | 0.9 |
| Ms | 7-10 | | 5-10 | 24 | 200-400 |
| $M_A$ | 7-11 | | 3-10 | 6 | 200-400 |

**Acknowledgement：**

We would like to thank the staff of the Shenguang II laser facility at Shanghai Institute of Optical and Fine Mechanics of Chinese Academy of Sciences for their help in carrying out these experiments. This work was funded by the Strategic Priority Research Program of Chinese Academy of Sciences (Grant No. XDB16000000, 41000000), the National Natural Science Foundation of China (Grant Nos. 12175230, 11775223, and 41804158), the Open Fund of the State Key Laboratory of High Field Laser Physics (SIOM), and the Fundamental Research Funds for the Central Universities.

**Author contributions:**

Guang-yue Hu and Quan-ming Lu conceived and led this project. Hui-bo Tang, Guang-yue Hu, Peng Hu, Yu-lin Wang, Zhen-chi Zhang, Xiang-bing Wang, Peng Yuan, Wei Liu, Chun-kai Yu, Jia-yi Zhao and Jin-can Wang participated in the design, setup, and execution of the experiment. Yu-fei Hao, Quan-ming Lu, Yu Zhang, Chuang Ren, and Ao Guo produced the theoretical and computational work. Hua-chong Si reconstructed the optical diagnostic results. Zhi-yong Xie prepared the target. Zhe Zhang, Xiao-hui Yuan, Da-wei Yuan, Jun Xiong, Zhi-heng Fang, Jian-cai Xu, Jing-Jing Ju, Guo-qiang Zhang, Jian-qiang Zhu, Ru-xin Li and Zhi-zhan Xu supplied diagnostic instruments and experimental facilities for the experiment.